\documentclass{article}
\usepackage{graphicx} 
\usepackage{cite}
\usepackage{amsmath,amssymb,amsfonts}
\usepackage{algorithmic}
\usepackage{textcomp}
\usepackage[hidelinks]{hyperref}

\usepackage{bm}
\usepackage{booktabs}
\makeatletter
\AtBeginDocument{\DeclareMathVersion{bold}
\SetSymbolFont{operators}{bold}{T1}{times}{b}{n}
\SetMathAlphabet{\mathrm}{bold}{T1}{times}{b}{n}
\SetMathAlphabet{\mathit}{bold}{T1}{times}{b}{it}
\SetMathAlphabet{\mathbf}{bold}{T1}{times}{b}{n}
\SetMathAlphabet{\mathtt}{bold}{OT1}{pcr}{b}{n}
\SetSymbolFont{symbols}{bold}{OMS}{cmsy}{b}{n}
\renewcommand\boldmath{\@nomath\boldmath\mathversion{bold}}}
\makeatother

\usepackage{wrapfig}

\title{Investigating the Impact of Rational Dilated Wavelet Transform on Motor Imagery EEG Decoding with Deep Learning Models}

\author{%
\centering
\begin{minipage}{0.95\textwidth}\centering
  Marco Siino$^{1,2}$\thanks{Corresponding author: \href{mailto:marco.siino@unict.it}{marco.siino@unict.it}, \href{mailto:marco.siino@unipa.it}{marco.siino@unipa.it}}\\
  Giuseppe Bonomo$^{2}$\\
  Rosario Sorbello$^{2}$\\
  Ilenia Tinnirello$^{2}$\\[0.8em]
  \small $^{1}$Department of Electrical, Electronic and Computer Engineering (DIEEI), University of Catania, 95123 Catania, Italy\\
  \small $^{2}$Department of Engineering, University of Palermo, 90128 Palermo, Italy
\end{minipage}
}
            
\date{October 2025}

\begin{document}

\maketitle


\begin{abstract}
The present study investigates the impact of the Rational Discrete Wavelet Transform (RDWT), used as a plug-in preprocessing step for motor imagery electroencephalographic (EEG) decoding prior to applying deep learning classifiers. A systematic paired evaluation (with/without RDWT) is conducted on four state-of-the-art deep learning architectures: EEGNet, ShallowConvNet, MBEEG\_SENet, and EEGTCNet. This evaluation was carried out across three benchmark datasets: High Gamma, BCI-IV-2a, and BCI-IV-2b. The performance of the RDWT is reported with subject-wise averages using accuracy and Cohen's kappa, complemented by subject-level analyses to identify when RDWT is beneficial. On BCI-IV-2a, RDWT yields clear average gains for EEGTCNet (+4.44 percentage points, pp; kappa +0.059) and MBEEG\_SENet (+2.23 pp; +0.030), with smaller improvements for EEGNet (+2.08 pp; +0.027) and ShallowConvNet (+0.58 pp; +0.008). On BCI-IV-2b, the enhancements observed are modest yet consistent for EEGNet (+0.21 pp; +0.044) and EEGTCNet (+0.28 pp; +0.077). On HGD, average effects are modest to positive, with the most significant gain observed for MBEEG\_SENet (+1.65 pp; +0.022), followed by EEGNet (+0.76 pp; +0.010) and EEGTCNet (+0.54 pp; +0.008). Inspection of the subject material reveals significant enhancements in challenging recordings (e.g., non-stationary sessions), indicating that RDWT can mitigate localized noise and enhance rhythm-specific information. In conclusion, RDWT is shown to be a low-overhead, architecture-aware preprocessing technique that can yield tangible gains in accuracy and agreement for deep model families and challenging subjects.
\end{abstract}

\section{Introduction}
\label{sec:introduction}
Decoding electroencephalographic (EEG) signals for Brain–Computer Interfaces (BCIs) remains challenging due to non-stationarity \cite{cao2011application,rasoulzadeh2017comparative}, low signal-to-noise ratio \cite{goldenholz2009mapping, hu2010novel}, and substantial inter-subject variability \cite{lo2009nonlinear, pfeffer2024exploring}. Variations in emotional and physiological states further inject noise and artifacts \cite{hosseini2010emotional, kim2013review}, making robust noise suppression and adaptive feature extraction critical prerequisites for reliable decoding \cite{Huber2021, Cong2023}.

Classical time–frequency techniques—Short-Time Fourier Transform (STFT) \cite{owens1988short, griffin1984signal} and standard Discrete Wavelet Transform (DWT) \cite{heil1989continuous}—are widely used for EEG preprocessing \cite{vidyasagar2024signal, gath2002tracking}, yet they face intrinsic trade-offs between temporal and spectral resolution. In particular, fixed integer dilation factors in conventional wavelet banks can limit adaptability to rapidly varying, non-stationary EEG dynamics. RDWT employs non-integer (rational) dilation factors (e.g., $3/2$, $5/3$), yielding a more flexible tiling of the time–frequency plane and potentially improving denoising and rhythm-specific enhancement in motor imagery settings. Rather than proposing a new network, our objective is to \emph{quantify if, when and how much} the usage of RDWT improve the performance across heterogeneous models and datasets. 

In this work we study, \emph{independently of any particular architecture}, the impact of a \emph{Rational Dilated Wavelet Transform} (RDWT) used as a plug-in preprocessing module before deep learning EEG classifiers. Unlike recent studies that explore the potential of RDWT for motor imagery (MI) signal classification using classical machine learning models \cite{alghayab2025eeg}, this work investigates the synergy between RDWT preprocessing and modern deep learning architectures. Specifically, we aim to determine if the beneficial effects of RDWT, such as mitigating noise and enhancing rhythm-specific information, translate into tangible performance gains when integrated into the preprocessing pipeline of state-of-the-art deep models. While classical classifiers have shown high accuracy with RDWT, their ability to capture complex non-linear features is limited compared to deep neural networks. By systematically evaluating the impact of RDWT on widely-used deep learning models like EEGNet, ShallowConvNet, MBEEG\_SENet, and EEGTCNet, we provide crucial insights into its utility as an architecture-agnostic preprocessing step for advanced EEG decoding. Our study addresses the question of whether RDWT should become a standard component in the pipeline for deep learning-based MI classification.

We conduct a systematic, paired comparison (with and without RDWT) on four state-of-the-art deep learning architectures - EEGNet, ShallowConvNet, MBEEG\_SENet and EEGTCNet on three benchmark dataset : High Gamma (HGD) \cite{schirrmeister2017deep}, BCI Competition IV-2a \cite{brunner2008bci}, and IV-2b \cite{leeb2008bci}. Performance is reported with subject-wise averages using accuracy and Cohen’s $\kappa$. Beyond aggregate metrics, we analyze subject-level heterogeneity to identify conditions under which RDWT is particularly beneficial (e.g., challenging subjects, non-stationary sessions, or backbones with strong temporal modeling).

\textbf{Preview of findings.} RDWT’s effect is \emph{architecture- and dataset-dependent}. On BCI-IV-2a, RDWT tends to improve models with explicit temporal stacks (e.g., TCN-based backbones), while effects on already multi-scale CNN pipelines are negligible to modest. On BCI-IV-2b, several subjects approach ceiling accuracy, yielding small average changes despite occasional subject-level gains. On HGD, average effects are modest but RDWT can materially help specific difficult subjects, indicating its value as a targeted denoising/representation step. The utilisation of RDWT is identified as a low-risk component that can yield practical benefits in the regimes where the additional time–frequency structure is most exploitable.

In summary, the paper reframes the process of EEG pre-processing as a design choice of paramount importance - often neglected in recent deep learning literature - and offers a study for the impact of RDWT in modern pipelines. Our contributions are:
\begin{itemize}
    \item A cross-architecture, cross-dataset evaluation of RDWT as a plug-in preprocessing step for EEG decoding before employing advanced deep learning architectures, with paired (with/without) comparisons on HGD, BCI-IV-2a, and BCI-IV-2b.
    \item A subject-wise analysis that reveals when RDWT is most beneficial (e.g., challenging subjects, non-stationary conditions, and temporal-modeling backbones), alongside aggregate accuracy and Cohen’s $\kappa$.
    \item Practical guidance for integrating RDWT in BCI pipelines, highlighting scenarios with expected payoff versus near-ceiling regimes where benefits are limited.
\end{itemize}

Taken together, our results position RDWT not as a universal remedy but as an \emph{architecture-aware}, data-dependent preprocessing tool that can improve robustness and accuracy in specific operating conditions, thereby supporting more reliable and generalizable EEG decoding also using deep learning models.

\section{Related Work}

\subsection{Time-Frequency Analysis in EEG}

Traditional signal processing techniques have long served as the foundation of EEG-based BCIs, offering a means to extract frequency-domain features from inherently non-stationary neural signals. The STFT is widely used to obtain time-localized spectral information via a sliding-window Fourier transform. However, its reliance on a fixed window size inherently limits the trade-off between time and frequency resolution \cite{vidyasagar2024signal}, making it less suited for tracking rapid transients in EEG.

Wavelet-based approaches, particularly the DWT, have demonstrated improved performance for EEG decoding by enabling multi-resolution analysis of signal components. DWT is particularly effective in isolating oscillatory activity across characteristic EEG frequency bands \cite{zhou2020wavelet}, and has been used both as a standalone preprocessing step and in conjunction with machine learning classifiers.

Recent advances have combined DWT with deep learning architectures, such as Convolutional Neural Networks (CNNs) and Recurrent Neural Networks (RNNs), where wavelet-decomposed EEG signals serve as inputs to enhance discriminative feature extraction \cite{bashivan2015learning}. Despite these improvements, both STFT and conventional DWT employ fixed or integer-valued dilation/scaling factors, which can fail to capture the variable temporal dynamics and scale-dependent structures present in EEG recordings.

While recent research has showcased the efficacy of RDWT in enhancing motor imagery (MI) signal classification, these efforts have predominantly relied on conventional machine learning models \cite{alghayab2025eeg}. The present study takes a distinct approach by bridging the gap between this powerful preprocessing technique and modern deep learning architectures. Instead of simply extracting features for traditional classifiers, our work systematically evaluates how RDWT integration impacts the performance of state-of-the-art deep models designed for EEG decoding. We aim to provide a comprehensive analysis of whether this approach can yield tangible improvements, thereby establishing a strong case for including RDWT as a standard preprocessing step in deep learning-based BCI pipelines.

The RDWT was introduced as a more flexible alternative \cite{bayram2009rdwt}. RDWT uses non-integer dilation factors, such as $3/2$ or $5/3$, to adapt the frequency resolution to the underlying signal structure more effectively. This provides a better time-frequency localization for non-stationary signals, making it particularly suitable for EEG preprocessing in BCI applications.

\subsection{Wavelet-Domain Preprocessing}

The input EEG signal is defined as $x_{e,c}(t)$ for event $e$, channel $c$, and discrete time index $t \in \{1, \dots, T\}$. To mitigate the effects of nonstationarity and enhance the signal representation, RatioWaveNet applies the RDWT as the first layer. The RDWT decomposes $x_{e,c}(t)$ into a set of subbands across multiple resolution levels $l = 1, \dots, L$, using rational dilation factors $d_l \in \mathbb{Q}$. The decomposition is performed as follows:

\begin{equation}
\begin{aligned}
a^{(l)}_{e,c}(t) &= \left(x^{(l-1)}_{e,c} * h^{(l)}_{\text{low}}\right)(t), \quad
d^{(l)}_{e,c}(t) = \left(x^{(l-1)}_{e,c} * h^{(l)}_{\text{high}}\right)(t),
\end{aligned}
\end{equation}

\noindent where $x^{(0)}_{e,c}(t) = x_{e,c}(t)$ and $*$ denotes the linear convolution. The filters $h^{(l)}_{\text{low}}$ and $h^{(l)}_{\text{high}}$ are rationally resampled versions of standard wavelet filters:

\begin{equation}
h^{(l)}(n) = \text{Resample}\left(h(n), \text{scale} = d_l\right),
\end{equation}

\noindent where $d_l \in \left\{\frac{3}{2}, \frac{5}{3}, \frac{7}{4}, \frac{9}{5}\right\}$. This multiresolution approach ensures translation invariance, which is critical for time-localized EEG features.

To suppress high-frequency noise, the detail coefficients are thresholded:

\begin{equation}
\tilde{d}^{(l)}_{e,c}(t) = 
\begin{cases}
d^{(l)}_{e,c}(t), & \text{if } |d^{(l)}_{e,c}(t)| \geq \tau, \\
0, & \text{otherwise},
\end{cases}
\end{equation}

\noindent where $\tau$ is a fixed amplitude threshold. The final signal is reconstructed by summing the final approximation and all thresholded detail components:

\begin{equation}
x^{(l)}_{e,c}(t) \leftarrow a^{(l)}_{e,c}(t), \quad
\hat{x}_{e,c}(t) = a^{(L)}_{e,c}(t) + \sum_{l=1}^{L} \tilde{d}^{(l)}_{e,c}(t).
\end{equation}

This preprocessing module ensures a denoised, translation-invariant multiresolution representation of the original EEG input, suitable for downstream deep learning models.

\section{Experimental Setup}

All experiments were conducted on a workstation equipped with an Intel\textregistered\ Xeon W7-3455 processor, an NVIDIA RTX 6000 Ada Generation GPU featuring 48 GB of memory, and 187 GiB of DDR5 system RAM. Training durations varied according to the dataset size and complexity: approximately 70 minutes for the High-Gamma Dataset, and around 42 minutes for both BCI-IV-2a and BCI-IV-2b datasets. Such variation reflects the inherent differences in data volume and complexity among the datasets. 

For all experiments, we adopted a consistent training protocol: optimization was performed using the Adam optimizer with a learning rate of 0.001, batch size of 64, and training proceeded for a maximum of 500 epochs, with early stopping triggered after 100 epochs without improvement.

Our evaluation was performed on three publicly available EEG benchmark datasets: BCI-IV-2a \cite{brunner2008bci}, BCI-IV-2b \cite{leeb2008bci}, and the High Gamma Dataset \cite{schirrmeister2017deep}. These datasets differ in acquisition hardware, experimental protocols, and number of subjects, providing a robust assessment of the model’s generalization capabilities. The code and experimental results are publicly available on GitHub.

\subsection{Dataset Description}

\label{sec:dataset_description}
In this section, we present the characteristics of the datasets utilized in our experiments.

\subsubsection{BCI IV 2a Dataset}

The BCI Competition IV Dataset 2a \cite{brunner2008bci} is a well-established resource for evaluating motor imagery  classification models. It includes EEG data collected from nine healthy volunteers (subjects A01–A09), each participating in two separate sessions recorded on different days to assess session variability. Participants were instructed to imagine the movement of one of four distinct body parts: left hand, right hand, both feet, or tongue. These actions were selected due to their distinct cortical representations, particularly in sensorimotor regions. Each session includes 288 randomized trials (72 per class). The experimental protocol followed a cue-based structure, starting with a fixation cross, followed by a directional arrow indicating the target motor imagery class. EEG signals were recorded via 22 Ag/AgCl electrodes following the 10–20 system, at a sampling rate of 250 Hz, and band-pass filtered between 0.5–100 Hz. Additionally, three EOG channels were recorded for ocular artifact correction. Signals were referenced to the left mastoid and grounded at AFz. Recordings were carried out in an electromagnetically shielded environment to minimize interference. This dataset is maintained by the Graz BCI Laboratory and has become a reference benchmark in motor imagery decoding, often used to test methods like CSP, deep learning models, and geometric classifiers.

\subsubsection{BCI IV 2b Dataset}

The BCI Competition IV Dataset 2b \cite{leeb2008bci} is another standard benchmark tailored for binary-class motor imagery studies, specifically focused on distinguishing between left- and right-hand imagery. EEG data were collected from nine healthy individuals (B01–B09) across five sessions. Sessions 1 and 4 were conducted without feedback, while the remaining sessions included real-time feedback. Each session consisted of 160 trials (80 per class), summing to 800 trials per subject. During each trial, subjects performed kinesthetic imagery of a hand movement in response to an arrow cue. EEG was recorded from three bipolar channels (C3, Cz, and C4) targeting the sensorimotor cortex, using a 250 Hz sampling rate and a band-pass of 0.5–100 Hz, with a 50 Hz notch filter to remove powerline noise. A minimal channel setup was used to emulate portable BCI scenarios. Feedback, when available, was presented as a bar indicating model predictions. Due to its simplicity and challenge, BCI-2B is frequently used to validate algorithms for motor imagery classification, including CSP variants and deep neural models.

\subsubsection{High-Gamma Dataset \cite{schirrmeister2017deep}}

The High-Gamma Dataset (HGD) \cite{schirrmeister2017deep} contains high-density EEG recordings from 14 healthy participants (6 female, 2 left-handed, mean age 27.2$\pm$3.6 years). Data were collected using 128 electrodes, with analysis focused on 44 channels covering the motor cortex. Subjects performed four tasks—imagined movements of the left hand, right hand, feet, and a resting state—over 13 runs, each comprising 80 trials of 4 seconds. On average, around 963 trials were recorded per subject, with the final two runs reserved for testing. Visual cues guided the tasks using directional arrows, with rest indicated by an upward arrow. The experimental setup emphasized minimal muscular movement, focusing on tasks like finger or toe tapping. Data acquisition was performed in a shielded environment with low-noise amplifiers and active shielding, and signals were recorded using the BCI2000 system at a sampling rate of 5 kHz \cite{schalk2004bci2000}. The study was ethically approved by the University of Freiburg and is widely used to benchmark deep learning-based EEG decoding techniques.

\subsection{Evaluation Metrics}

The performance of the preprocessing in the models is assessed using two standard metrics: accuracy and the Cohen's Kappa coefficient. These metrics are widely adopted in the EEG-based classification literature \cite{kappa, huang2020intelligent, metrics} for benchmarking purposes. Accuracy is defined as the proportion of correctly classified instances over the total number of samples. Cohen’s Kappa quantifies the agreement between predicted and true labels while accounting for the agreement occurring by chance. 

The definitions of Accuracy and Kappa are given by Equation~\eqref{accuracy kappa}:

\begin{equation}
\label{accuracy kappa}
\text{Accuracy} = \frac{1}{n} \sum_{i=1}^{n} \frac{TP_i}{I_i}, \quad
\text{Kappa} = \frac{1}{n} \sum_{a=1}^{n} \frac{P_a - P_e}{1 - P_e}
\end{equation}

where $TP_i$ denotes the number of correctly predicted samples for class $i$, $I_i$ is the total number of samples in class $i$, $n$ is the total number of classes, $P_a$ is the observed accuracy across classes, and $P_e$ is the expected accuracy under random chance.

\subsection{Baseline Comparison}

For comparative evaluation, we selected a set of state-of-the-art models commonly employed in EEG classification tasks: \textit{EEGNet}~\cite{lawhern2018eegnet}, \textit{ShallowConvNet}~\cite{Wang2019shallow}, \textit{MBEEG\_SENet}~\cite{Altuwaijri2022multi}, and \textit{EEG-TCNet}~\cite{ingolfsson2020eeg}. Each baseline was trained using the original hyperparameters reported in their respective works. A standardized preprocessing, training, and evaluation pipeline was employed across all models to ensure fairness in comparison.

Briefly, \textit{EEGNet} utilizes a combination of 2D temporal, depthwise, and separable convolutions tailored for general BCI classification tasks. \textit{ShallowConvNet} comprises two convolutional layers followed by mean pooling operations optimized for EEG decoding. \textit{MBEEG\_SENet} incorporates a multi-branch convolutional architecture with squeeze-and-excitation blocks to capture multi-scale EEG features. \textit{EEG-TCNet} combines CNNs and TCNs to effectively model temporal dependencies.

\section{Results}
\label{sec:results}

This section presents a comprehensive, paired evaluation of the RDWT as a plug-in preprocessing step across multiple EEG classifiers. For each models (EEGNet, ShallowConvNet, MBEEG\_SENet, EEGTCNet) and dataset (BCI-IV-2a, BCI-IV-2b, HGD), we report subject-wise averages of \emph{classification accuracy} and \emph{Cohen’s~$\kappa$} for two conditions: \textit{None} (no preprocessing) and \textit{RDWT}. 

Our goal is to isolate the contribution of RDWT independently of architectural specifics: by comparing each model to its own baseline on the same participants, we directly assess whether RDWT systematically improves agreement and accuracy. Beyond aggregate statistics, we also examine subject-level variability to identify regimes where RDWT is most effective (e.g., recordings with pronounced non-stationarity or localized noise) and where gains are naturally limited (e.g., near-ceiling performance). The following subsections detail these results per dataset, highlighting consistent patterns of improvement and noting cases in which RDWT yields negligible or mixed changes, thereby providing an evidence-based view of when this preprocessing is most beneficial.

\subsubsection{Effect of RDWT Preprocessing on Model Performance (BCI-IV-2a)}

Table~\ref{tab:bci2a_rdwt-none} reports a paired comparison (with/without RDWT) across four baseline architectures on BCI-IV-2a. On average, RDWT yields \emph{consistent} improvements for all models considered, although the magnitude is clearly architecture-dependent. EEGTCNet exhibits the largest gain in average accuracy (+4.44\,pp; 64.35\%\,$\rightarrow$\,68.79\%) together with the largest increase in Cohen’s $\kappa$ (+0.059). MBEEG\_SENet follows with a +2.23\,pp gain (70.49\%\,$\rightarrow$\,72.72\%; $\kappa$ +0.030), while EEGNet improves by +2.08\,pp (68.02\%\,$\rightarrow$\,70.10\%; $\kappa$ +0.027). ShallowConvNet shows a smaller but positive shift (+0.58\,pp; 65.74\%\,$\rightarrow$\,66.32\%; $\kappa$ +0.008). 

A subject-level inspection reveals heterogeneous effects that clarify where RDWT is most beneficial. For EEGTCNet, gains are pronounced on S01 (+12.16\,pp) and S04 (+10.07\,pp), with additional improvements on S03 (+4.51\,pp), S05 (+4.86\,pp), and S06 (+3.47\,pp), and only a marginal decrease on S07 (–0.34\,pp). MBEEG\_SENet shows a marked jump on S05 (+22.22\,pp) and smaller increases on S09 (+2.43\,pp), with modest declines on S03/S04 (–1.04/–1.38\,pp) and S07 (–3.13\,pp). EEGNet benefits notably on S09 (+10.07\,pp), S06 (+3.82\,pp), and S04 (+3.47\,pp), while dropping on S05 (–2.78\,pp) and S08 (–2.08\,pp). ShallowConvNet exhibits a large improvement on S04 (+12.50\,pp), but decreases on S05/S06/S07 (–6.25/–2.43/–4.16\,pp), yielding a modest net gain overall.

Taken together, these results indicate that RDWT is most impactful when the backbone can exploit enhanced time–frequency structure - particularly in temporally expressive models such as TCN-style stacks (EEGTCNet) - and for subjects whose recordings present greater non-stationarity or localized noise. Conversely, compact CNNs with limited temporal modeling capacity tend to realize smaller average benefits, with mixed subject-wise responses. In practical terms, RDWT emerges as a low-overhead, architecture-aware preprocessing step that can yield meaningful accuracy and agreement improvements on BCI-IV-2a, especially for models and subjects that can capitalize on its multi-resolution representation.

\begin{table*}[h]
  \caption{Comparison of different methods with and without the RDWT module on BCI-IV-2a. As shown in bold all the averaged metrics are improved by the use of RDWT.}
  \label{tab:bci2a_rdwt-none}
  \centering
  \resizebox{\textwidth}{!}{%
  \begin{tabular}{llcccccccccccl}
    \toprule
    \textbf{Model} & \textbf{Preproc.} & S01 & S02 & S03 & S04 & S05 & S06 & S07 & S08 & S09 & \textbf{Avg.} & \textbf{Kappa} \\
    \midrule
    EEGNet          & None  & 77.08 & 51.74 & 91.67 & 54.17 & 63.89 & 45.83 & 83.68 & 77.08 & 67.01 & 68.02 & 0.5740 \\
                    & RDWT  & 79.86 & 51.74 & 92.71 & 57.64 & 61.11 & 49.65 & 86.11 & 75.00 & 77.08 & \textbf{70.10} & \textbf{0.6010} \\
    \midrule
    ShallowConvNet  & None  & 72.92 & 46.88 & 83.68 & 47.92 & 63.19 & 40.97 & 81.94 & 78.82 & 75.35 & 65.74 & 0.5430 \\
                    & RDWT  & 74.31 & 48.26 & 83.33 & 60.42 & 56.94 & 38.54 & 77.78 & 79.86 & 77.43 & \textbf{66.32} & \textbf{0.5510} \\
    \midrule
    MBEEG\_SENet    & None  & 82.29 & 53.82 & 92.36 & 65.62 & 44.79 & 57.64 & 86.46 & 79.51 & 71.88 & 70.49 & 0.6060 \\
                    & RDWT  & 82.64 & 54.17 & 91.32 & 64.24 & 67.01 & 57.99 & 83.33 & 79.51 & 74.31 & \textbf{72.72} & \textbf{0.6360} \\
    \midrule
    EEGTCNet        & None  & 63.19 & 49.65 & 81.25 & 50.69 & 62.50 & 46.18 & 80.90 & 68.75 & 76.04 & 64.35 & 0.5250 \\
                    & RDWT  & 75.35 & 52.78 & 85.76 & 60.76 & 67.36 & 49.65 & 80.56 & 70.14 & 76.74 & \textbf{68.79} & \textbf{0.5840} \\
    \bottomrule
  \end{tabular}
  }
\end{table*}

\subsubsection{Effect of RDWT Preprocessing on Model Performance (BCI-IV-2b)}

Table~\ref{tab:bci2b_rdwt-none} indicates that the impact of RDWT on BCI-IV-2b is modest due to the effect of ceiling performance on several subjects ($\geq\!99\%$). In average accuracy, RDWT yields small but consistent gains for EEGNet (+0.21\,pp; 95.85\%\,$\rightarrow$\,96.06\%) and EEGTCNet (+0.28\,pp; 95.81\%\,$\rightarrow$\,96.09\%), and a minor improvement for ShallowConvNet (+0.09\,pp; 95.85\%\,$\rightarrow$\,95.94\%). MBEEG\_SENet shows a slight increase in average accuracy (+0.14\,pp; 96.39\%\,$\rightarrow$\,96.53\%) but a decrease in agreement ($\kappa$: 0.671\,$\rightarrow$\,0.646, –0.025). Agreement gains are otherwise positive for EEGNet ($\kappa$ +0.044), EEGTCNet ($\kappa$ +0.077), and ShallowConvNet ($\kappa$ +0.005). 

Subject-wise patterns reveal where RDWT is most beneficial. For EEGNet, S09 exhibits a marked improvement (+5.62\,pp; 93.51\%\,$\rightarrow$\,99.13\%), with S02 and S06 reaching 100\%, while S04 decreases (–4.44\,pp). EEGTCNet gains on S01 (+1.42\,pp), S03 (+1.09\,pp), S05 (+1.29\,pp), S08 (+1.45\,pp), and S09 (+1.30\,pp), but drops on S04 (–2.22\,pp) and marginally on S06/S07. ShallowConvNet benefits chiefly on S04 (+4.45\,pp) amid small declines on S03 (–1.09\,pp), S07 (–1.86\,pp), and S08 (–0.73\,pp). MBEEG\_SENet shows mixed changes—improvements on S02 (+0.90\,pp) and S04 (+2.22\,pp) contrasted with decreases on S03 (–1.45\,pp) and S09 (–0.43\,pp)—which plausibly explains the slight rise in accuracy alongside a lower $\kappa$.

Overall, RDWT can deliver small accuracy and agreement gains for backbones able to exploit additional time–frequency structure (notably TCN-based EEGTCNet and, to a lesser extent, EEGNet). In near-ceiling regimes typical of BCI-IV-2b, its average benefits are limited and architecture-specific, with occasional subject-level improvements offset by minor degradations on others.

\begin{table*}[h]
  \caption{Comparison of different methods with and without the RDWT module on BCI-IV-2b. As shown in bold all the averaged metrics are improved by the use of RDWT.}
  \label{tab:bci2b_rdwt-none}
  \centering
  \resizebox{\textwidth}{!}{%
  \begin{tabular}{llccccccccccc}
    \toprule
    \textbf{Model} & \textbf{Preproc.} & S01 & S02 & S03 & S04 & S05 & S06 & S07 & S08 & S09 & \textbf{Avg.} & \textbf{Kappa} \\
    \midrule
    EEGNet          & None  & 98.58 & 99.10 & 98.19 & 84.44 & 92.95 & 99.53 & 97.41 & 98.91 & 93.51 & 95.85 & 0.5960 \\
                    & RDWT  & 99.29 & {100.00} & 98.55 & 80.00 & 92.31 & {100.00} & 97.04 & 98.19 & 99.13 & \textbf{96.06} & \textbf{0.6400} \\
    \midrule
    ShallowConvNet  & None  & 99.65 & 99.10 & 97.83 & 82.22 & 91.67 & 99.06 & 95.19 & 99.28 & 98.70 & 95.85 & 0.6180 \\
                    & RDWT  & 99.65 & 99.10 & 96.74 & 86.67 & 91.67 & 99.06 & 93.33 & 98.55 & 98.70 & \textbf{95.94} & \textbf{0.6230} \\
    \midrule
    MBEEG\_SENet    & None  & 98.94 & 98.20 & 99.28 & 82.22 & 92.95 & {100.00} & 97.04 & 98.91 & 100.00 & 96.39 & 0.6710 \\
                    & RDWT  & 98.94 & 99.10 & 97.83 & 84.44 & 92.95 & {100.00} & 97.04 & 98.91 & 99.57 & \textbf{96.53} & \textbf{0.6460} \\
    \midrule
    EEGTCNet        & None  & 98.23 & 99.10 & 97.10 & 84.44 & 90.38 & {100.00} & 97.41 & 98.19 & 97.40 & 95.81 & 0.5890 \\
                    & RDWT  & 99.65 & 98.20 & 98.19 & 82.22 & 91.67 & 99.53 & 97.04 & {99.64} & 98.70 & \textbf{96.09} & \textbf{0.6660} \\
    \bottomrule
  \end{tabular}
  }
\end{table*}

\subsubsection{Effect of RDWT Preprocessing on Model Performance (HGD Dataset)}

Table~\ref{tab:hgd_none} shows that RDWT produces consistent average gains also on HGD, with clear architecture-dependent effects. In terms of average accuracy, the largest improvement is observed for MBEEG\_SENet (+1.65\,pp; 88.61\%\,$\rightarrow$\,90.26\%), accompanied by a higher agreement ($\kappa$: 0.848\,$\rightarrow$\,0.870, +0.022). EEGNet and EEGTCNet also benefit, with accuracy shifts of +0.76\,pp (87.32\%\,$\rightarrow$\,88.08\%; $\kappa$ +0.010) and +0.54\,pp (86.60\%\,$\rightarrow$\,87.14\%; $\kappa$ +0.008), respectively. ShallowConvNet exhibits a near-null change (+0.22\,pp; 87.05\%\,$\rightarrow$\,87.27\%; $\kappa$ +0.003). 

Subject-level analyses reveal that RDWT can be markedly beneficial on difficult cases. The most prominent example is S14, where accuracy increases for EEGNet (+21.25\,pp; 57.50\%\,$\rightarrow$\,78.75\%), MBEEG\_SENet (+21.26\,pp; 60.62\%\,$\rightarrow$\,81.88\%), and ShallowConvNet (+6.26\,pp). On S11, EEGTCNet improves by +7.50\,pp, whereas MBEEG\_SENet remains unchanged. On S6, RDWT benefits EEGTCNet (+3.13\,pp) and MBEEG\_SENet (+1.26\,pp). These results suggest that RDWT is most effective when additional time–frequency structure aids temporal modeling or mitigates non-stationarity; conversely, when recordings are already clean or the backbone captures rich multi-scale features, net improvements are limited.

In absolute terms, the best overall configuration on HGD is MBEEG\_SENet with RDWT (90.26\% average accuracy; $\kappa=0.870$).

\begin{table*}[ht]
\caption{Comparison of different methods with and without the RDWT module on HGD (14 subjects). As shown in bold all the averaged metrics are improved by the use of RDWT.}
\label{tab:hgd_none}
\centering
\tiny
\setlength{\tabcolsep}{2pt}
\resizebox{\textwidth}{!}{%
\begin{tabular}{llcccccccccccccccc}
\toprule
\textbf{Model} & \textbf{Preproc.} & S1 & S2 & S3 & S4 & S5 & S6 & S7 & S8 & S9 & S10 & S11 & S12 & S13 & S14 & \textbf{Avg.} & \textbf{Kappa} \\
\midrule
EEGNet          & None & 89.38 & 81.25 & 97.50 & 95.62 & 90.00 & 94.38 & 91.19 & 87.50 & 95.62 & 86.25 & 76.88 & 93.12 & 86.25 & 57.50 & {87.32} & {0.8310} \\
                & RDWT & 87.50 & 83.12 & 93.75 & 93.12 & 90.62 & 88.12 & 90.57 & 90.00 & 95.62 & 88.12 & 71.88 & 91.88 & 90.00 & 78.75 & \textbf{88.08} & \textbf{0.8410} \\
\midrule
ShallowConvNet  & None & 88.75 & 87.50 & 96.25 & 95.62 & 88.75 & 91.88 & 88.05 & 85.62 & 95.62 & 85.62 & 68.75 & 91.88 & 76.25 & 78.12 & 87.05 & 0.8270 \\
                & RDWT & 86.88 & 90.00 & {98.75} & 96.88 & 90.62 & 90.62 & 86.79 & 80.00 & 96.25 & 89.38 & 61.25 & 93.75 & 76.25 & 84.38 & \textbf{87.27} & \textbf{0.8300} \\
\midrule
MBEEG\_SENet    & None & 92.50 & 84.38 & 97.50 & 96.25 & 91.25 & 90.62 & 91.82 & 89.38 & 95.62 & 90.62 & 76.88 & 93.75 & 89.38 & 60.62 & 88.61 & 0.8480 \\
                & RDWT & 91.88 & 91.88 & 95.62 & 94.38 & {94.38} & 91.88 & 91.19 & 92.50 & 95.00 & 86.25 & 76.88 & 91.88 & 88.12 & 81.88 & \textbf{90.26} &\textbf{ 0.8700} \\
\midrule
EEGTCNet        & None & 87.50 & 90.62 & 95.62 & 90.62 & 90.00 & 85.62 & 93.08 & 87.50 & 95.00 & 85.00 & 65.62 & 93.12 & 84.38 & 68.75 & 86.60 & 0.8210 \\
                & RDWT & 86.25 & 90.62 & 94.38 & 92.50 & 90.00 & 88.75 & 88.68 & 88.12 & 95.00 & 88.12 & 73.12 & 90.62 & 90.62 & 63.12 & \textbf{87.14} & \textbf{0.8290} \\
\bottomrule
\end{tabular}
}
\end{table*}

\subsection{Scalogram-Based Analysis of EEG Signals}

\begin{figure}[!t]
    \centering
    \includegraphics[width=\columnwidth]{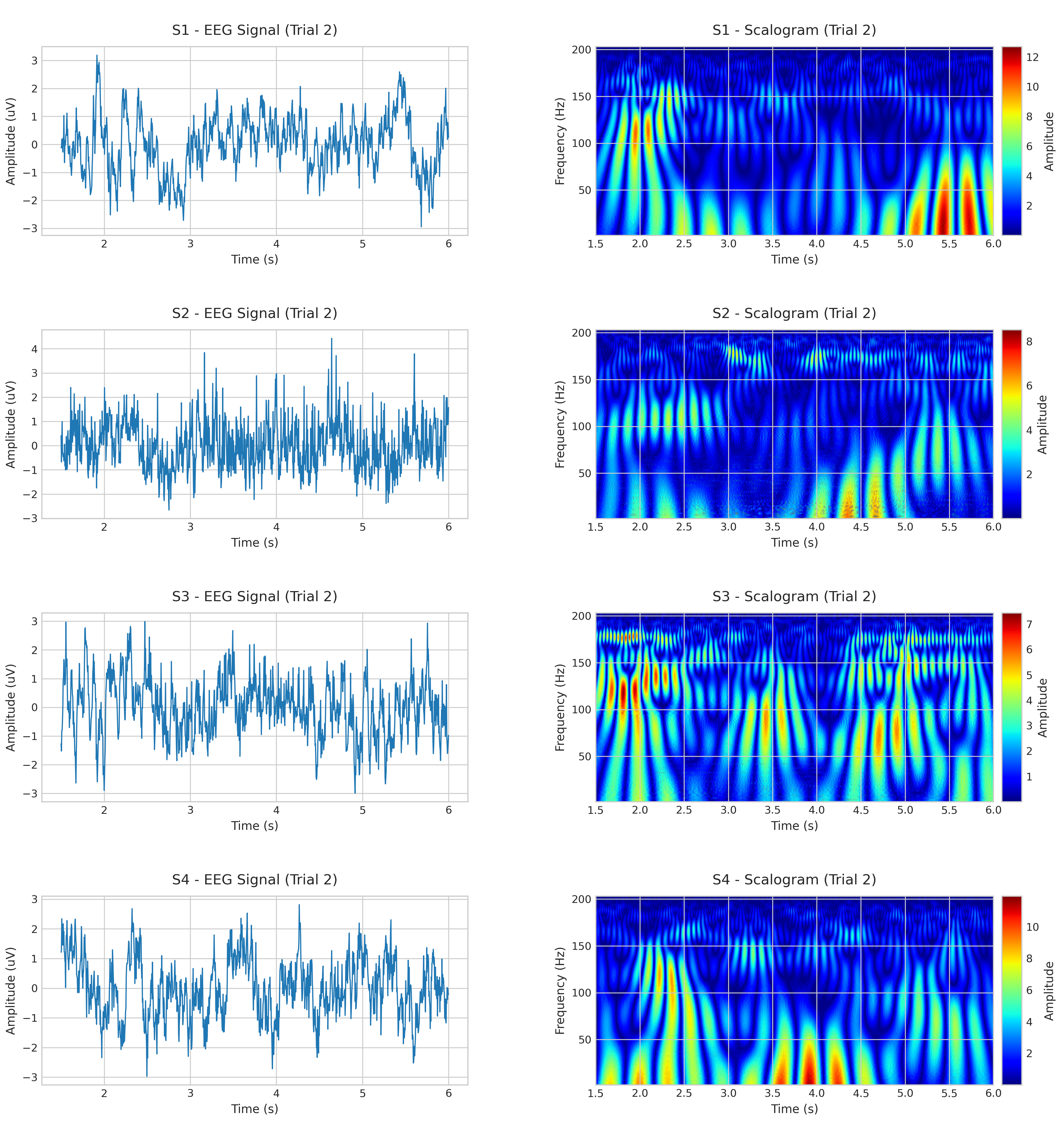}
    \caption{Comparison of EEG signals and scalograms (Left Hand Movement, Cz, Trial 2) for subjects S1–S4. Left: time-domain signals; right: time-frequency maps. The spectral differences reflect inter-subject variability in motor-related activity.}
    \label{fig:eeg_scalogram_comparison}
\end{figure}

Fig.~\ref{fig:eeg_scalogram_comparison} presents a comparative visualization of EEG signals and their corresponding scalograms for the "Left Hand Movement" class, recorded from the Cz channel during the second trial across four subjects (S1–S4). The left panels show raw EEG signals in the time domain, while the right panels depict their time-frequency representations obtained using the continuous wavelet transform (CWT). While the EEG amplitudes are similar in range, the scalograms reveal subject-specific spectral patterns, particularly in the 10–30 Hz and 100–150 Hz bands. These differences may reflect inter-subject variability in motor intention processing and support the use of time-frequency features for improving classification performance in motor imagery tasks.

\begin{figure}
    \centering
    \includegraphics[width=1.0\linewidth]{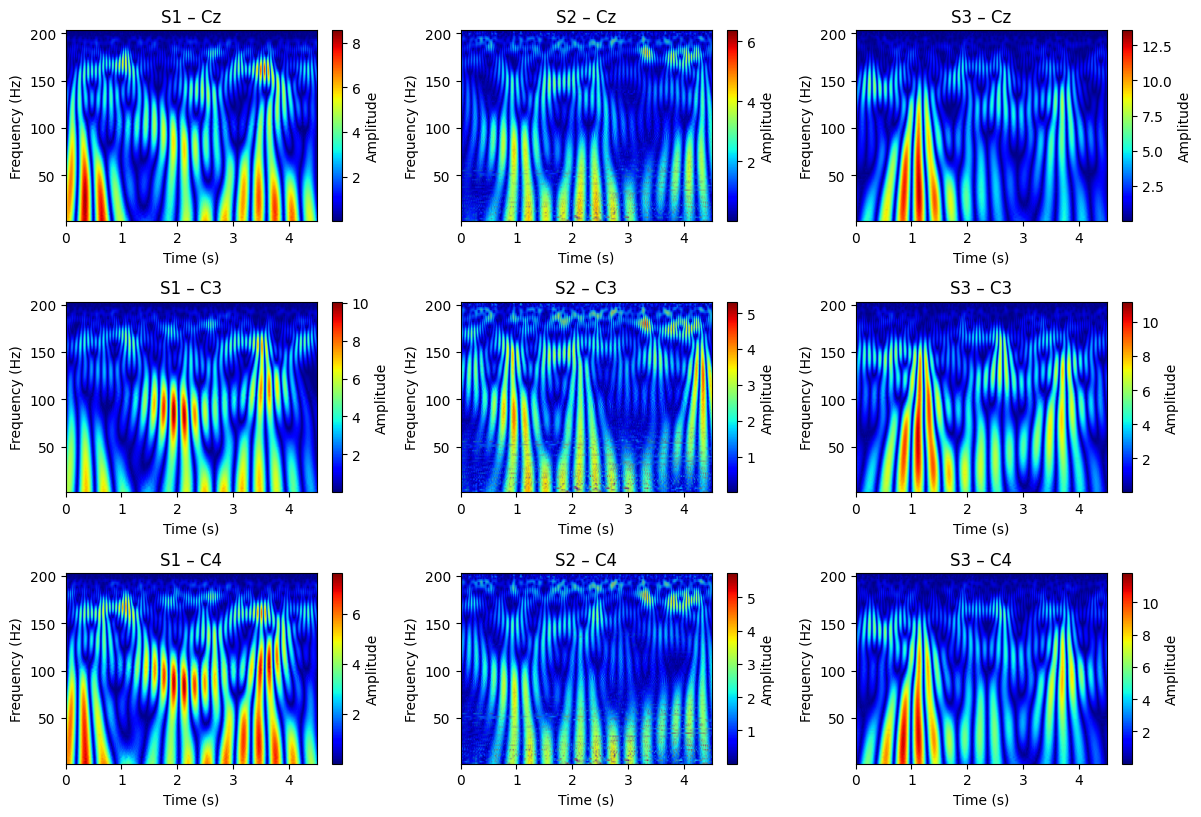}
    \caption{Scalograms for S1–S3 across channels Cz, C3, C4 during left-hand motor imagery. The high spectral similarity across channels suggests redundancy, supporting the choice of a single representative channel for further analysis.}
   
    \label{fig:scalogram_subjects_channels}
\end{figure}

Time–frequency representations offer an intuitive tool for visual inspection of EEG signals in motor imagery tasks. However, a detailed scalogram-based analysis revealed that frequency-domain patterns alone do not provide sufficient discriminative information across classes or subjects. As illustrated in Fig.~\ref{fig:scalogram_subjects_channels}, for three representative subjects and across three different EEG channels (Cz, C3, and C4), the scalograms display highly similar spectral structures, regardless of the channel considered. This suggests a high level of redundancy in the frequency response across channels, which justifies our decision to retain a single representative channel in the subsequent analysis.

\begin{figure*}
    \centering
    \includegraphics[width=1\linewidth]{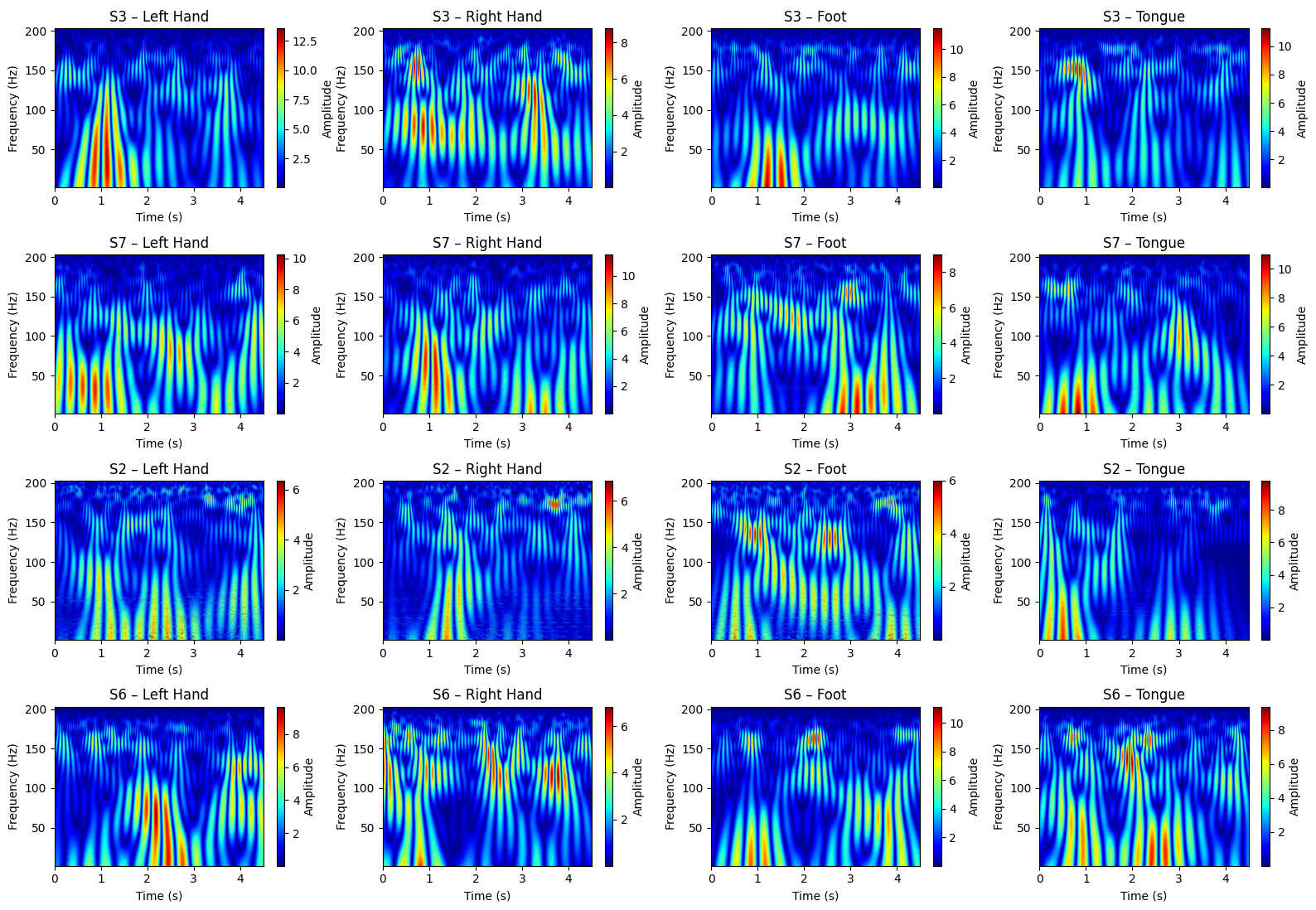}
      \caption{Scalogram visualization of the first trial for each motor imagery class (left hand, right hand, foot, tongue) in four BCI-IV-2a subjects: S3 and S7 (best classification) vs. S2 and S6 (worst). Rows: subjects; columns: classes. The absence of consistent patterns highlights the limited discriminative power of frequency-domain features.}
    \label{fig:scalogram_subjects_classes}
\end{figure*}

Furthermore, Fig.~\ref{fig:scalogram_subjects_classes} compares the spectral responses of the first four subjects of the BCI Competition IV-2a dataset while imagining the four movements (same channel). The Figure is a visualization of the first trial for each motor imagery class (left hand, right hand, foot, tongue) across four subjects (S3, S7, S2 and S6) from the BCI-IV-2a dataset. The first two (i.e., S3 and S7) are the subjects providing the highest classification accuracies while the last two (i.e., S2 and S6) are the ones with the worst classification accuracies. Consistently with the previous observation, the spectrograms do not exhibit distinctive or consistent patterns, either within or across subjects and classes. This visual evidence aligns with the relatively low classification performance obtained using frequency-domain features, supporting the conclusion that purely spectral information is insufficient for reliable discrimination in this context.

Our observations, corroborated by recent studies \cite{altaheri2023deep,li2019novel}, suggest that frequency-based analyses may fail to capture the dynamic temporal variations inherent in EEG data, leading to suboptimal classification performance. For example, in \cite{li2019novel}, the authors employed CNNs to investigate the impact of generating topological representations from motor imagery EEG signals in both the time and frequency domains. Their findings indicate that maps derived from the time domain significantly outperform those constructed using frequency-domain information, highlighting the superiority of time-resolved features for capturing relevant neural dynamics in motor imagery tasks. Finally, As reported in \cite{altaheri2023deep}, studies utilizing raw EEG signal inputs on the most frequently used motor imagery dataset (BCI Competition IV-2a) achieved an average accuracy of 77.65\%. This performance slightly surpasses that of approaches based on extracted features (77.02\%), spectral images (76.80\%), and topological maps (76.5\%). Similarly, for the second most commonly employed dataset (BCI Competition IV-2b), models trained directly on raw EEG signals attained an average accuracy of 83.22\%, outperforming those relying on extracted features (81.56\%) and spectral image inputs (82.07\%). Consequently, these results advocate for the adoption of time-resolved and adaptive signal processing methods that better accommodate the non-stationary nature of EEG signals in brain-computer interface applications.

\begin{figure*}[t]
    \centering
    \includegraphics[width=\textwidth]{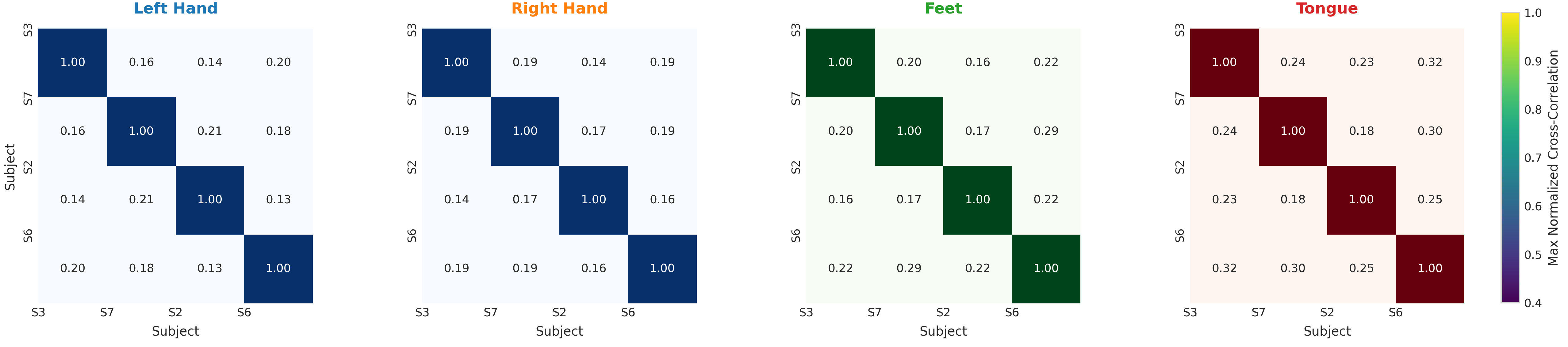}
    \caption{Inter-subject correlation matrices (Trial 3, Cz) for four motor imagery tasks. Matrices show pairwise maximum normalized cross-correlations between S3, S7, S2, and S6. Higher off-diagonal values indicate greater cross-subject similarity.}
    \label{fig:inter_subject_corr}
\end{figure*}

Figure~\ref{fig:inter_subject_corr} illustrates the inter-subject correlation matrices for four distinct motor imagery tasks—left hand, right hand, feet, and tongue—based on the maximum normalized cross-correlation values computed from EEG signals at channel Cz during Trial 2. Also in this case the correlation matrices are related to the two top-classification-accuracies subjects (i.e., S3 and S7)  and to the two worst-classification-accuracies subjects (i.e., S2 and S6). A key observation from these heatmaps is the consistently high self-correlation (1.00) for each subject across all tasks, as expected. More critically, the figure highlights the varying degrees of cross-subject correlation. Notably, the correlations between subjects S2 and S6 are consistently lower compared to the correlations involving S3 and S7, regardless of the motor imagery task. For instance, in the "Left Hand" task, the cross-correlation between S2 and S6 is 0.13, significantly lower than the 0.20 correlation between S3 and S7 for the same task. This pervasive pattern of lower cross-subject correlations for S2 and S6 across all tasks suggests a reduced consistency in their brain activity patterns compared to S3 and S7. This finding is particularly pertinent when considering the previously discussed challenges in classifying data from S2 and S6 (as highlighted in the context of Figure 6), as it directly indicates a lower degree of shared neural features that a classification algorithm could exploit across these more challenging subjects. The overall implication is that the lack of robust inter-subject commonalities, especially for subjects like S2 and S6, poses a significant hurdle for achieving high classification accuracies in brain-computer interface systems that rely on generalized models or feature extraction techniques.

\section{Conclusion}

This study assessed the impact of RDWT preprocessing on EEG decoding in a paired (with/without) setting across four widely used backbones—EEGNet, ShallowConvNet, MBEEG\_SENet, and EEGTCNet—on three benchmark motor-imagery datasets (BCI-IV-2a, BCI-IV-2b, HGD). Our analysis isolates the contribution of RDWT as a plug-in, architecture-agnostic step placed upstream of otherwise unchanged classifiers.

\textbf{Summary of findings.} RDWT’s effect is unequivocally \emph{architecture- and dataset-dependent}.  
On BCI-IV-2a (Table~\ref{tab:bci2a_rdwt-none}), RDWT yields clear average gains for EEGTCNet (+4.44\,pp; $\kappa$ +0.059) and MBEEG\_SENet (+2.23\,pp; $\kappa$ +0.030), with smaller but positive shifts for EEGNet (+2.08\,pp; $\kappa$ +0.027) and ShallowConvNet (+0.58\,pp; $\kappa$ +0.008).  
On BCI-IV-2b (Table~\ref{tab:bci2b_rdwt-none}), where several subjects operate near ceiling, improvements are necessarily modest: EEGTCNet (+0.28\,pp; $\kappa$ +0.077) and EEGNet (+0.21\,pp; $\kappa$ +0.044) benefit slightly; ShallowConvNet shows a near-null change (+0.09\,pp; $\kappa$ +0.005); MBEEG\_SENet exhibits a small accuracy increase (+0.14\,pp) alongside a lower agreement (–0.025 $\kappa$).  
On HGD (Table~\ref{tab:hgd_none}), average effects are modest-to-positive across all four models, with MBEEG\_SENet showing the largest gain (+1.65\,pp; $\kappa$ +0.022), followed by EEGNet (+0.76\,pp; $\kappa$ +0.010), EEGTCNet (+0.54\,pp; $\kappa$ +0.008), and ShallowConvNet (+0.22\,pp; $\kappa$ +0.003). The present tables do not report $p$-values; statistical significance was therefore not assessed here.

\textbf{When and why RDWT helps.} Subject-wise patterns reveal that RDWT can be decisively beneficial on difficult cases (e.g., S14 in HGD, with double-digit gains for multiple backbones), suggesting that the added time–frequency structure mitigates non-stationarity and improves rhythm isolation. Architectures with stronger temporal modeling capacity (e.g., TCN-style stacks such as EEGTCNet) tend to leverage RDWT more consistently, while compact CNNs display smaller, heterogeneous responses that depend on subject and dataset headroom.

\textbf{Practical guidance.} RDWT is a low-overhead, plug-in preprocessing suitable for pipelines that (i) face pronounced inter/intra-subject variability, (ii) rely on temporal modeling to capture task-relevant dynamics, and/or (iii) are not already operating at ceiling. In near-ceiling regimes (typical of BCI-IV-2b), expected gains are small; selective or subject-adaptive activation of RDWT may be preferable to blanket application.

\textbf{Limitations and future work.} Average improvements, while consistent in direction for several model–dataset pairs, are modest; future studies should increase cohort sizes and report paired statistical testing. Automating RDWT hyperparameters (rational dilation factors, mother wavelet choice) via data-driven or differentiable selection, as well as learning subject- or session-adaptive policies that trigger RDWT only when predicted to help, are promising directions. Extending the analysis beyond motor imagery will further clarify the scope of RDWT in broader EEG paradigms.

\textbf{Conclusion.} RDWT preprocessing is not a universal remedy, but it is a practical and sometimes impactful addition that can improve robustness and agreement for specific backbones and challenging subjects. As such, it merits consideration as an optional, architecture-aware component in modern EEG deep learning-based decoding pipelines.

\section*{Author Contributions}

M.~Siino led the conceptualization of the study, designed the methodology, developed the software, and conducted the investigation. He also contributed to the original draft writing, revision, supervision, and project administration. G.~Bonomo contributed to software development, data curation, validation, investigation, and writing of the original draft, including result visualization. R.~Sorbello assisted with methodology design, validation, supervision, and contributed to both drafting and reviewing the manuscript. I.~Tinnirello provided resources, supervised the project, contributed to methodology discussions, manuscript review, and secured funding.



\begin{thebibliography}{10}
\providecommand{\url}[1]{#1}
\csname url@samestyle\endcsname
\providecommand{\newblock}{\relax}
\providecommand{\bibinfo}[2]{#2}
\providecommand{\BIBentrySTDinterwordspacing}{\spaceskip=0pt\relax}
\providecommand{\BIBentryALTinterwordstretchfactor}{4}
\providecommand{\BIBentryALTinterwordspacing}{\spaceskip=\fontdimen2\font plus
\BIBentryALTinterwordstretchfactor\fontdimen3\font minus \fontdimen4\font\relax}
\providecommand{\BIBforeignlanguage}[2]{{%
\expandafter\ifx\csname l@#1\endcsname\relax
\typeout{** WARNING: IEEEtran.bst: No hyphenation pattern has been}%
\typeout{** loaded for the language `#1'. Using the pattern for}%
\typeout{** the default language instead.}%
\else
\language=\csname l@#1\endcsname
\fi
#2}}
\providecommand{\BIBdecl}{\relax}
\BIBdecl

\bibitem{cao2011application}
C.~Cao and S.~Slobounov, ``Application of a novel measure of eeg non-stationarity as ‘shannon-entropy of the peak frequency shifting’for detecting residual abnormalities in concussed individuals,'' \emph{Clinical Neurophysiology}, vol. 122, no.~7, pp. 1314--1321, 2011.

\bibitem{rasoulzadeh2017comparative}
V.~Rasoulzadeh, E.~Erkus, T.~Yogurt, I.~Ulusoy, and S.~A. Zergero{\u{g}}lu, ``A comparative stationarity analysis of eeg signals,'' \emph{Annals of Operations Research}, vol. 258, pp. 133--157, 2017.

\bibitem{goldenholz2009mapping}
D.~M. Goldenholz, S.~P. Ahlfors, M.~S. H{\"a}m{\"a}l{\"a}inen, D.~Sharon, M.~Ishitobi, L.~M. Vaina, and S.~M. Stufflebeam, ``Mapping the signal-to-noise-ratios of cortical sources in magnetoencephalography and electroencephalography,'' \emph{Human brain mapping}, vol.~30, no.~4, pp. 1077--1086, 2009.

\bibitem{hu2010novel}
L.~Hu, A.~Mouraux, Y.~Hu, and G.~D. Iannetti, ``A novel approach for enhancing the signal-to-noise ratio and detecting automatically event-related potentials (erps) in single trials,'' \emph{Neuroimage}, vol.~50, no.~1, pp. 99--111, 2010.

\bibitem{lo2009nonlinear}
M.-T. Lo, P.-H. Tsai, P.-F. Lin, C.~Lin, and Y.~L. Hsin, ``The nonlinear and nonstationary properties in eeg signals: probing the complex fluctuations by hilbert--huang transform,'' \emph{Advances in Adaptive Data Analysis}, vol.~1, no.~03, pp. 461--482, 2009.

\bibitem{pfeffer2024exploring}
M.~A. Pfeffer, S.~S.~H. Ling, and J.~K.~W. Wong, ``Exploring the frontier: Transformer-based models in {EEG} signal analysis for brain-computer interfaces,'' \emph{Computers in Biology and Medicine}, p. 108705, 2024.

\bibitem{hosseini2010emotional}
S.~A. Hosseini and M.~A. Khalilzadeh, ``Emotional stress recognition system using eeg and psychophysiological signals: Using new labelling process of eeg signals in emotional stress state,'' in \emph{2010 international conference on biomedical engineering and computer science}.\hskip 1em plus 0.5em minus 0.4em\relax IEEE, 2010, pp. 1--6.

\bibitem{kim2013review}
M.-K. Kim, M.~Kim, E.~Oh, and S.-P. Kim, ``A review on the computational methods for emotional state estimation from the human eeg,'' \emph{Computational and mathematical methods in medicine}, vol. 2013, no.~1, p. 573734, 2013.

\bibitem{Huber2021}
N.~R. Huber, A.~D. Missert, H.~Gong, and B.~E. Nett, ``Random search as a neural network optimization strategy for convolutional-neural-network (cnn)-based noise reduction in ct,'' in \emph{Medical Imaging 2021: Image Processing}, vol. 11596.\hskip 1em plus 0.5em minus 0.4em\relax SPIE, Feb 2021, pp. 509--515.

\bibitem{Cong2023}
S.~Cong and Y.~Zhou, ``A review of convolutional neural network architectures and their optimizations,'' \emph{Artificial Intelligence Review}, vol.~56, no.~3, pp. 1905--1969, Jun 2023.

\bibitem{owens1988short}
F.~Owens and M.~Murphy, ``A short-time fourier transform,'' \emph{Signal Processing}, vol.~14, no.~1, pp. 3--10, 1988.

\bibitem{griffin1984signal}
D.~Griffin and J.~Lim, ``Signal estimation from modified short-time fourier transform,'' \emph{IEEE Transactions on acoustics, speech, and signal processing}, vol.~32, no.~2, pp. 236--243, 1984.

\bibitem{heil1989continuous}
C.~E. Heil and D.~F. Walnut, ``Continuous and discrete wavelet transforms,'' \emph{SIAM review}, vol.~31, no.~4, pp. 628--666, 1989.

\bibitem{vidyasagar2024signal}
K.~E. Ch~Vidyasagar, K.~Revanth~Kumar, G.~N.~K. Anantha~Sai, M.~Ruchita, and M.~J. Saikia, ``Signal to image conversion and convolutional neural networks for physiological signal processing: A review,'' \emph{IEEE Access}, vol.~12, pp. 66\,726--66\,764, 2024.

\bibitem{gath2002tracking}
I.~Gath, C.~Feuerstein, D.~T. Pham, and G.~Rondouin, ``On the tracking of rapid dynamic changes in seizure eeg,'' \emph{IEEE Transactions on Biomedical Engineering}, vol.~39, no.~9, pp. 952--958, 2002.

\bibitem{alghayab2025eeg}
H.~R. {Al Ghayab}, Y.~Li, M.~Diykh, A.~Sahi, S.~Abdulla, and A.~R. Alkhuwaylidee, ``Eeg based over-complete rational dilation wavelet transform coupled with autoregressive for motor imagery classification,'' \emph{Expert Systems with Applications}, vol. 269, p. 126433, 2025.

\bibitem{schirrmeister2017deep}
R.~T. Schirrmeister, J.~T. Springenberg, L.~D.~J. Fiederer, M.~Glasstetter, K.~Eggensperger, M.~Tangermann, F.~Hutter, W.~Burgard, and T.~Ball, ``Deep learning with convolutional neural networks for eeg decoding and visualization,'' \emph{Human brain mapping}, vol.~38, no.~11, pp. 5391--5420, 2017.

\bibitem{brunner2008bci}
C.~Brunner, R.~Leeb, G.~M{\"u}ller-Putz, A.~Schl{\"o}gl, and G.~Pfurtscheller, ``Bci competition 2008--graz data set a,'' \emph{Institute for knowledge discovery (laboratory of brain-computer interfaces), Graz University of Technology}, vol.~16, no. 1-6, p.~1, 2008.

\bibitem{leeb2008bci}
R.~Leeb, C.~Brunner, G.~M{\"u}ller-Putz, A.~Schl{\"o}gl, and G.~Pfurtscheller, ``Bci competition 2008--graz data set b,'' \emph{Graz University of Technology, Austria}, vol.~16, pp. 1--6, 2008.

\bibitem{zhou2020wavelet}
Y.~Zhou, S.~Wang, Z.~Chen, H.~Yang, and L.~Li, ``Wavelet transform time–frequency image and convolutional neural network for motor imagery eeg classification,'' \emph{Sensors}, vol.~20, no.~9, p. 2514, 2020.

\bibitem{bashivan2015learning}
P.~Bashivan, I.~Rish, M.~Yeasin, and N.~Codella, ``Learning representations from eeg with deep recurrent-convolutional neural networks,'' in \emph{International Conference on Learning Representations (ICLR)}, 2016.

\bibitem{bayram2009rdwt}
I.~Bayram and I.~W. Selesnick, ``Frequency-domain design of overcomplete rational-dilation wavelet transforms,'' \emph{IEEE Transactions on Signal Processing}, vol.~57, no.~8, pp. 2957--2972, 2009.

\bibitem{schalk2004bci2000}
G.~Schalk, D.~J. McFarland, T.~Hinterberger, N.~Birbaumer, and J.~R. Wolpaw, ``Bci2000: A general-purpose brain-computer interface (bci) system,'' in \emph{IEEE Transactions on biomedical engineering}, vol.~51.\hskip 1em plus 0.5em minus 0.4em\relax IEEE, 2004, pp. 1034--1043.

\bibitem{kappa}
\BIBentryALTinterwordspacing
J.~Cohen, ``A coefficient of agreement for nominal scales,'' \emph{Educational and Psychological Measurement}, vol.~20, no.~1, pp. 37--46, 1960. [Online]. Available: \url{https://doi.org/10.1177/001316446002000104}
\BIBentrySTDinterwordspacing

\bibitem{huang2020intelligent}
\BIBentryALTinterwordspacing
X.~Huang, Y.~Li, Y.~Chen, Y.~Lin, and D.~Yao, ``An intelligent eeg classification methodology based on sparse representation enhanced deep learning networks,'' \emph{Frontiers in Neuroscience}, vol.~14, p. 808, 2020. [Online]. Available: \url{https://www.frontiersin.org/articles/10.3389/fnins.2020.00808/full}
\BIBentrySTDinterwordspacing

\bibitem{metrics}
\BIBentryALTinterwordspacing
J.~Opitz, ``A closer look at classification evaluation metrics and a critical reflection of common evaluation practice,'' \emph{Transactions of the Association for Computational Linguistics}, vol.~12, pp. 820--836, 06 2024. [Online]. Available: \url{https://doi.org/10.1162/tacl\_a\_00675}
\BIBentrySTDinterwordspacing

\bibitem{lawhern2018eegnet}
V.~J. Lawhern, A.~J. Solon, N.~R. Waytowich, C.~Gordon, C.~Hung, and B.~J. Lance, ``Eegnet: A compact convolutional neural network for eeg-based brain–computer interfaces,'' \emph{Journal of neural engineering}, vol.~15, no.~5, p. 056013, 2018.

\bibitem{Wang2019shallow}
T.~Wang, E.~Dong, S.~Du, and C.~Jia, ``A shallow convolutional neural network for classifying mi-eeg,'' in \emph{2019 Chinese Automation Congress (CAC)}, Nov 2019, pp. 5837--5841.

\bibitem{Altuwaijri2022multi}
G.~A. Altuwaijri, G.~Muhammad, H.~Altaheri, and M.~Alsulaiman, ``A multi-branch convolutional neural network with squeeze-and-excitation attention blocks for eeg-based motor imagery signals classification,'' \emph{Diagnostics}, vol.~12, no.~4, p. 995, Apr 2022.

\bibitem{ingolfsson2020eeg}
T.~M. Ingolfsson, M.~Hersche, X.~Wang, N.~Kobayashi, L.~Cavigelli, and L.~Benini, ``{EEG-TCNet}: An accurate temporal convolutional network for embedded motor-imagery brain–machine interfaces,'' in \emph{2020 IEEE International Conference on Systems, Man, and Cybernetics (SMC)}, 2020, pp. 2958--2965.

\bibitem{altaheri2023deep}
H.~Altaheri, G.~Muhammad, M.~Alsulaiman, S.~U. Amin, G.~A. Altuwaijri, W.~Abdul, M.~A. Bencherif, and M.~Faisal, ``Deep learning techniques for classification of electroencephalogram (eeg) motor imagery (mi) signals: A review,'' \emph{Neural Computing and Applications}, vol.~35, no.~20, pp. 14\,681--14\,722, 2023.

\bibitem{li2019novel}
M.-A. Li, J.-F. Han, and L.-J. Duan, ``A novel mi-eeg imaging with the location information of electrodes,'' \emph{IEEE Access}, vol.~8, pp. 3197--3211, 2019.

\end{thebibliography}
\end{document}